# System Test of the ATLAS Muon Spectrometer in the H8 Beam at the CERN SPS

Erez Etzion

On behalf of the ATLAS Muon Collaboration

*Abstract*--An extensive system test of the ATLAS muon spectrometer has been performed in the H8 beam line at the CERN SPS during the last four years. This spectrometer will use pressurized Monitored Drift Tube (MDT) chambers and Cathode Strip Chambers (CSC) for precision tracking, Resistive Plate Chambers (RPCs) for triggering in the barrel and Thin Gap Chambers (TGCs) for triggering in the end-cap region. The test set-up emulates one projective tower of the barrel (six MDT chambers and six RPCs) and one end-cap octant (six MDT chambers, A CSC and three TGCs). The barrel and end-cap stands have also been equipped with optical alignment systems, aiming at a relative positioning of the precision chambers in each tower to 30-40 micrometers.

In addition to the performance of the detectors and the alignment scheme, many other systems aspects of the ATLAS muon spectrometer have been tested and validated with this setup, such as the mechanical detector integration and installation, the detector control system, the data acquisition, high level trigger software and off-line event reconstruction. Measurements with muon energies ranging from 20 to 300 GeV have allowed measuring the trigger and tracking performance of this set-up, in a configuration very similar to the final spectrometer.

A special bunched muon beam with 25 ns bunch spacing, emulating the LHC bunch structure, has been used to study the timing resolution and bunch identification performance of the trigger chambers. The ATLAS first-level trigger chain has been operated with muon trigger signals for the first time.

Manuscript received October, 26, 2004. This work is supported by the Israel Science Foundation and the German Israeli Foundation.
E. Etzion is with the School of Physics and Astronomy, Raymond and Beverly Sackler Faculty of Excat Sciences, Tel Aviv University, Tel Aviv 69978, Israel (telephone: +41764870750, e-mail: erez.etzion@cern.ch).

## I. INTRODUCTION

THE Large Hadron Collider (LHC) currently under construction at the European Center for Nuclear Research (CERN) is designed to collide proton-proton beams at the energy of 7 TeV, the highest energy ever reached in a particle accelerator. The ATLAS detector built for the LHC experiment will collect events at a design luminosity of $10^{34}$ $cm^{-2}s^{-1}$ in a proton bunch crossing rate of 40 MHz (3-20 collisions every 25 ns).

The largest part in the ATLAS detector is the Muon spectrometer [1]. It has been designed to provide a standalone trigger on single muons with transverse momentum of several GeV as well as to measure final state muons with a momentum resolution of about 3% over most of the expected momentum range; a resolution of 10% is expected at transverse momenta of 1 TeV. The ATLAS Muon spectrometer is a $4\pi$ detector consists of four types of detector technologies.

Over most of the spectrometer acceptance, Monitored Drift Tube (MDT) chambers are used for the precision measurement of muon tracks. The MDTs are built from aluminum tubes filled with Ar-$CO_2$ at a pressure of 3 bars. Operated at a gas gain of $2 \times 10^4$, each tube measures charged particle tracks with an average spatial resolution better than 80 micrometers.

Cathode Strip Chambers (CSC) are used for muon momentum measurement in the inner part of the end-cap regions where the background rate is expected to reach 1 KHz/$cm^2$.

Three stations of Resistive Plate Chambers (RPC) are employed for triggering muons in the barrel region where three Thin Gap Chambers (TGC) stations serve the same purpose in the higher background region of the end-cap. Two RPC stations are attached to the MDT middle station and provide the low-Pt (transverse momentum) trigger. Similar functionality is assigned to the two TGC double gap units (Doublets) installed close to the end-cap MDT middle station. For the high-Pt muon trigger the low-Pt signal is combined with the one provided by the RPCs attached to the barrel MDT outer station or the three gaps TGC units (Triplet) in the end-cap region. The signal from the trigger chambers are amplified discriminated and digitally shaped on the detector and sent to an ASIC-based coincidence matrix board. These devices perform the functions needed for the trigger algorithm and apply the Pt cuts following preset thresholds. The trigger chambers are also used to provide the coordinate along the drift tubes that can not be measured by the MDT chambers. For that purpose several additional TGC chambers are also installed close to the end-cap inner MDT part to improve the measurement precision of this coordinate.

A large scale test stand of the ATLAS detector including all the Muon spectrometer components has been operated in the CERN north area on the H8 beam line since the year 2000. A beam of up to 320 GeV provided by the SPS accelerator was used to study different aspects of the spectrometer. In this paper we summarize the performance of the Muon spectrometer and its different components during the 2004 H8 runs.

## II. THE H8 MUON SETUP

The H8 Muon stand consists of two parts: a barrel stand and an end-cap stand. The barrel setup is reproducing one ATLAS barrel sector with its MDT and RPC stations. It consists of six MDT chambers (two of each type: inner, middle and outer chambers) fully instrumented with Front End electronics (FE) read with a Muon Readout Driver (MROD) and fully equipped with an alignment system. There are six RPC doublets: four middle chambers (BML) and two outer chambers (BOL) in the barrel set-up. Two additional barrel stations where used in the test stand: one inner chamber (BIL) on a rotating support for calibration studies and one outer (BOS) station (MDT+RPC) upstream of the muon wall for noise and Combined Test Beam (CTB) studies.

In the end-cap stand which reproduces a muon spectrometer end-cap sector there are 11 MDT chambers: two inner (EI), two middle (EM) and two outer (EO). As in the barrel they are fully instrumented with FE and read out through one MROD. The chambers are equipped with the complete alignment system and calibrated sensors for absolute alignment. One CSC chamber has recently being installed and should be integrated soon in the combined data taking. For triggering there are three TGC units: one triplet and two doublets fully instrumented with on-chamber electronics.

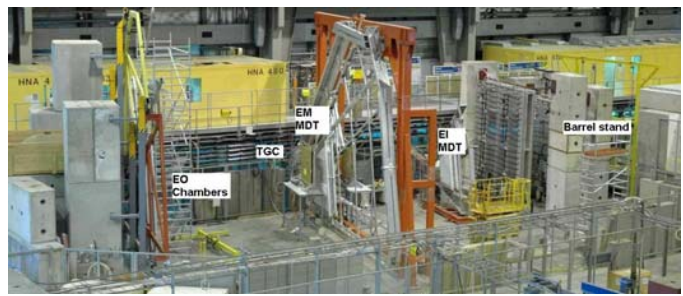

Figure 1 - A picture of the Muon spectrometer section setup in the H8 area. On the right handside the barrel three stations consisting of two inner, two middle and two outer MDTs. Attached to the middle and outer barrel MDT are the partner RPCs. On the left the end-cap chambers with six MDTs, two doublets and one TGC triplet.

## III. THE DETECTOR SLOW CONTROL, ONLINE SOFTWARE AND DATABASE

The monitoring and operation on crucial parameters of the spectrometer working conditions are controlled via the Detector Control System (DCS) [2]. The monitoring is done via a commercial control environment, PVSS-II, which serves all the LHC experiments. The readout, monitoring and analysis initialization is distributed via DIM, a data transfer system [3]. The MDTs initialization and configuration, the temperature monitoring, the control on the FE low voltage (LV) power supplies and the barrel and end-cap alignment system were controlled under the frame of the MDT DCS.

The TGC DCS system [4] was used to control the CAEN LV and HV power supplies by the PVSS process communicating via OPC server [5]. The DCS was used to set the readout threshold. The DCS was running a chamber charge monitoring embedded in the DCS boards. The DCS board provided monitoring of the analog charge of a wire in a chamber over some time interval. The resulted histograms by the data analysis performed in the ELMB [6] (see

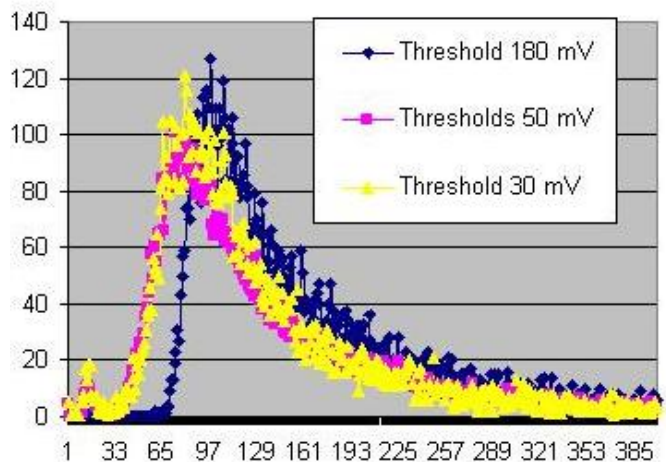

Figure 2) were distributed through the CAN bus network [7].

The figure shows how the system allowed controlling the level of noise and trigger loss by monitoring histograms at the PVSS workstation and setting the threshold in a chamber. TGC-DCS communicated directly with the TGC DAQ and used direct SQL queries to store and retrieve information from/to the conditions/configurations database (DB).

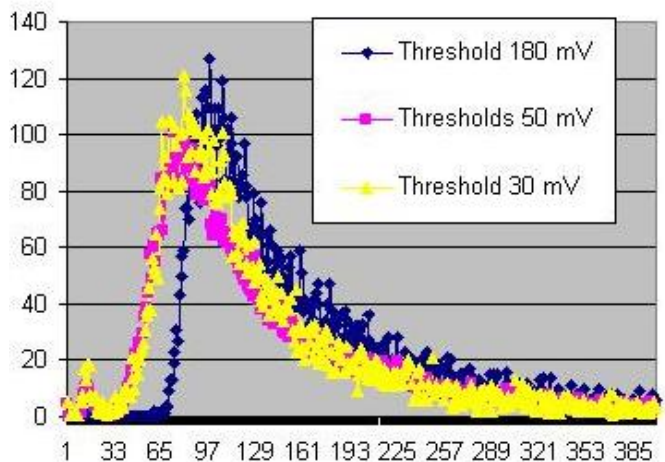

Figure 2 - Monitoring histogram of one analog TGC channel read via the DCS. The blue histogram is collected with a readout threshold of 180 mV and the pink and yellow are collected with a threshold of 50 mV and 30 mV respectively.

RPC DCS which recently moved to PVSS was used to monitor and control the conditions of: the gas status (flux of components), manifold pressure, HV status (monitoring the voltage and current), RPCs gap current and LV supplying the FE electronics.

We exploit the test beam date to test the ATLAS online software. Two software packages were utilized for on-line monitoring. One is the monitoring framework GNAM which was co-developed with the ATLAS Tile Calorimeter and was successfully used in the test beam [8]. The other option which was tested is monitoring within ATHENA, the ATLAS offline software framework [9].

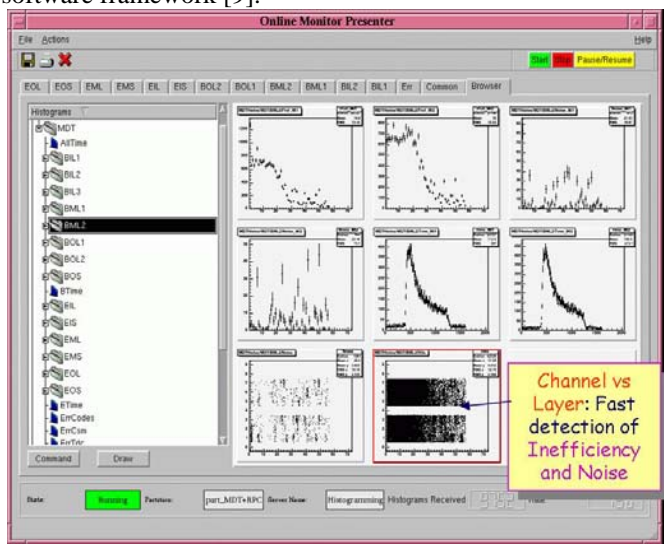

Figure 3 - A window of the online monitoring as used in the test beam.

The muon spectrometer used a MySQL DB server serving the non-event data storage. Nearly a complete loop of applications using the so called conditions DB has been implemented. All the quantities needed were stored and read out. High rate access for raw alignment image results was used (every ~2 sec). Access from ATHENA is currently under development, first prototype has been implemented for alignment corrections.

Two approaches of communication to the DB were implemented and tested. One way was connection to the DB via a dedicated API and the second method was a direct SQL connection from DAQ and the DCS systems. The second end-up with a full relational DB functionality as required by operation and maintenance of several component of the ATLAS detector.

## IV. ALIGNMENT SYSTEM, OFFLINE SOFTWARE AND SIMULATION STUDIES

In order to fully exploit the intrinsic resolution of the tracking system, the signal wires must be positioned with an accuracy of 20 micrometers with respect to the chamber reference system. The precision of the mechanical construction is monitored with elaborate X-ray techniques to an accuracy of 5 µm. Optical alignment systems have been developed which allow to measure on-line internal deformations of chambers under gravity, and the relative alignment of chambers traversed by the same muon track, to a typical accuracy of 30-40 µm [10]. For that purpose a three dimensional grid of alignment devices consists of over more than 10,000 sensors optically connect groups of chambers together. In case of the end-cap it is also connected to the alignment bars. These bars which are up to 9.6 m long are nested in each layer of chambers. They are self-aligning units measuring their own positions and therefore can be considered precision rulers.

The performance of the barrel and the end-cap alignment system has been validated in the H8 test beam performing controlled physical movements and rotations of the chambers or by inducing thermal expansion of the support structure. The barrel and the end-cap alignment systems have been able to track changes in muon sagita's to an accuracy of about 15 microns under normal temperature variation and controlled movements. The system was found stable over a period of months. Figure 4 demonstrates how the alignment correct controlled movements as demonstrated with the sagita plotted for the nominal setup and for controlled movement on the left and with the alignment correction applied on the right. The plots clearly show that once the alignment is taken into account the three distributions perfectly match.

In parallel to some stand alone code written specifically to analyze the test beam results, the ATHENA environment was used as the offline monitoring, reconstruction and analysis tool for the test beam data. Many aspects of the ATLAS offline environment were used such as data converters reading the raw data from all the Muon spectrometer technologies (MDT, CSC, RPC, TGC, MUCTPI [11]) following the scheme designed for the ATLAS.

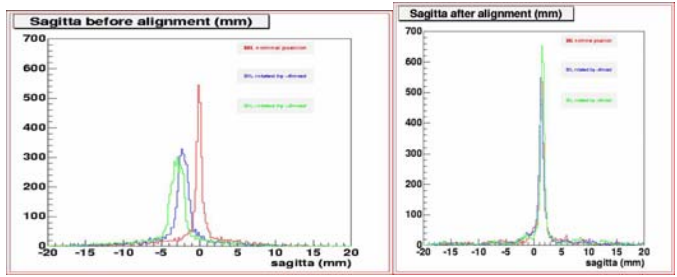

Figure 4 - The two figures show the distribution of the sagita plotted on the left before alignment and on the right after alignment. The three different lines in the two plots correspond to a plot without movement compared with controlled movements of six and eight mrad. On the left before alignment and on the right after alignment is applied.

Offline software was used for the three stages: second level trigger, event filter and the offline analysis code. Similar strategy for detector description and data analysis were successfully implemented in the test beam as in the preparation for the ATLAS data taking stage (the Data Challenge studies). The combined test beam enabled combined tests of the Muon spectrometer with other parts of ATLAS as well as among the Muon sub-detectors. In Figure 5 depicted as an example the correlation between projection in the horizontal plane of tracks in the Muon system and the inner detector. A good agreement ($p1 = 0.99 \pm 0.08$) is demonstrated in the plot while the shift of 80 mm resulted from a known shift from the original position of that size.

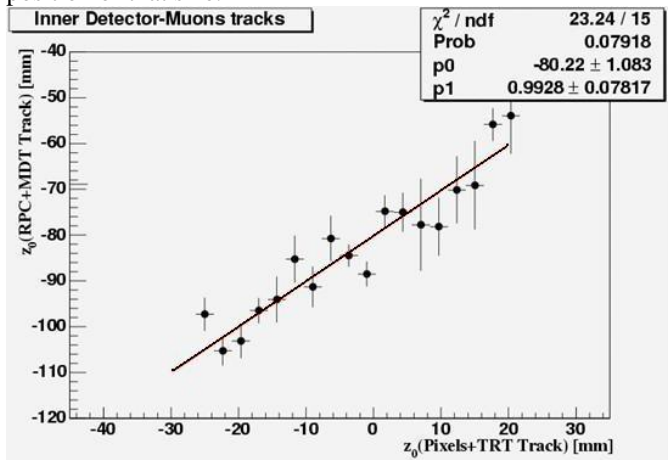

Figure 5 - The correlation between the Muon system tracking and the tracking in the inner detector (Pixels and TRT tracking).

The simulation of muons interacting with the H8 chambers and the material has been performed with Geant4 (version 6) [12] running within the ATHENA framework. The Geometry was taken from an external ASCII database following the same format that is used for the ATLAS Muon system description. The comparison of reconstructed real data and simulated events can provide at this point an important feedback to the validation of both reconstruction and simulation software. A comparison between the energy resolutions measured in the simulation and in the data showed a similar dependence on the actual test-beam beam energy (or simulated energy in the case simulated event).

## V. TRIGGER CHAMBER PERFORMANCE AND BEAM ENERGY MEASUREMENT

One of the main goals of this test was to study the trigger and read-out of the trigger chambers. A performance of the trigger chamber was studied: efficiency of wires and strips of the TGC doublets and triplets in low and high Pt cuts, efficiency of the RPCs as a function of the applied voltage, measurement of the cluster sizes, and uniformity of the response. Another goal set for the tests was to study the readout electronics and the trigger using final electronics and cabling schemes for the signals going from the FE electronics.
The special bunched muon beam with 25 ns bunch spacing, emulating the LHC bunch structure, has been used to study the timing resolution and bunch identification of the trigger chamber. The ATLAS first level trigger chain [13] has been operated with the muon trigger signals for the first time.

The TGC had some upgrades and modification since the trigger was successfully tested in the 2003 test beam. The TGC took data during the run with 25 ns bunched beam. It provided a validation of the design of the end-cap Muon Level-1 trigger scheme. The ATLAS Central Trigger Processor (CTP) receives trigger information from the calorimeter and Muon trigger processors and generates the "Level-1 Accept" based on a trigger menu. The Muon Trigger to CTP Interface (MUCTPI) is used to connect the TGC and RPC information to the CTP. During the 2004 25 ns bunched beam, TGC Level-1 trigger signals were sent through the MUCTPI and a comparison between the TGC sector logic output and the MUCTPI matched perfectly. TGC successfully tested the integration with the condition DB. The TGC performed successfully with the results on Level-1 trigger of 99.4% efficiency for low-Pt and 98.1% for high-Pt. Figure 6 demonstrates the correlation between track position as derived from the MDT tracking and the groups of wires hit in the TGC. This can be interpreted as a position correlation between the two. On the second plot a breakdown of the efficiency given separately for wires and strips, doublets and triplets. All the values are above 99.5%.

During 2004 test beam RPC electronic was tested with the final cabling scheme for the signals going from the front-end to the read-out electronics. The performance of the RPC chambers was studied: efficiency as a function of the applied voltage, measurement of cluster sizes and uniformity of response.

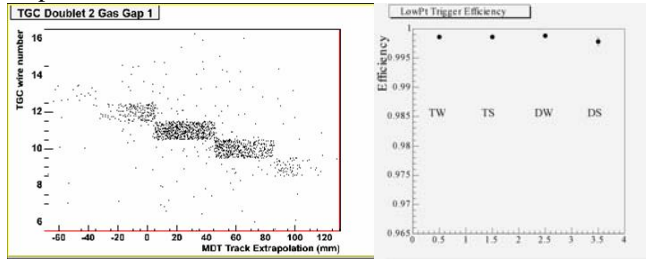

Figure 6 - On the left a correlation between the MDT track extrapolated to the TGC position and the TGC wire group number, on the right the efficiency derived for the TGC Triples wires (TW), Triplet strips (TS) Doublet wires (DW) and doublet strips (DS).

Figure 7 presents the correlation between the positions as derived from the MDTs and the one calculated from the RPC strip hit position. The second plot shows the dependence of the BML RPC efficiency as a function of the voltage for different threshold values. One reaches efficiencies about 95% when the voltage is above 9.4kV.

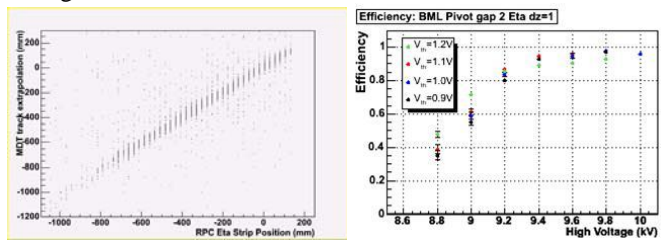

Figure 7 - On left a correlation between the RPC eta strip position and the extrapolated position from the the MDT tracks, on the right the efficiency derived for BML RPC chamber as a function of the HV supplied to the chamber for several values of readout threshold.

## VI. SUMMARY

A large scale test stand of the ATLAS Muon spectrometer has been operated in the H8 test beam at the CERN SPS. The setup consisted of a barrel stand reproducing one barrel setup and a stand reproducing an end-cap sector both fully instrumented with electronics readout and complete alignment systems. A muon beam with momenta ranging from 20 to 320 GeV was used to study and integrate many aspects of the Muon spectrometer. Special runs with 25 ns bunch spacing were dedicated for trigger timing resolution and bunch identification studies. The spectrometer has been extensively tested and validated with this setup. The studies covered: mechanical detector integration, integration between the spectrometer different technologies (MDT, CSC, RPC and TGC), different subsystems (readout, DCS, alignment) different spectrometer tasks (trigger, tracking) and different software tools (data acquisition, databases, high level trigger software, on-line and offline monitoring and reconstruction, alignment and calibration). Some of the measurements of quantities such as track reconstruction, energy resolution, trigger performance, alignment and comparison with simulation are presented in the paper. Other than that the test beam has provided the ATLAS Muon community with a unique opportunity to run a significant part of the detector in a long scale experiment. It gave a field to test the on-line and offline simulation and analysis tools and to have a comparison of the simulation to real data results.

## VII. ACKNOWLEDGMENT

The work reported here represents a joint effort of many individuals in the ATLAS Muon collaboration and in the ATLAS Trigger Data Acquisition community. We would like to thank them all. A special thank goes to Stefano Rosati for his assistance in preparing this summary. The support of the CERN staff operating the SPS and the H8 beam line is gratefully acknowledged. We thank the funding agencies for the financial support.